\documentclass[final,5p,twocolumn]{elsarticle}
\pdfoutput=1
\usepackage[utf8]{inputenc}
\usepackage{epsf}
\usepackage{epsfig}
\usepackage{graphics}
\usepackage{graphicx}
\usepackage{amssymb}
\usepackage{amsmath}
\usepackage{multirow}
\usepackage{color}
\usepackage{xcolor}
\usepackage{ulem}

\newcommand{\nc}{\newcommand}
\nc{\postscript}[2] 
{\setlength{\epsfxsize}{#2\hsize}\centerline{\epsfbox{#1}}}
\nc{\non}{\nonumber}
\nc{\hc}{\hbox {h.c.}} \nc{\re}{\hbox {Re}} 
\nc{\mev}{\hbox {MeV}} \nc{\gev}{\;\hbox {GeV}} \nc{\tev}{\;\hbox {TeV}}
\def\lsim{\mathrel{\raise.3ex\hbox{$<$\kern-.75em\lower1ex\hbox{$\sim$}}}}
\def\gsim{\mathrel{\raise.3ex\hbox{$>$\kern-.75em\lower1ex\hbox{$\sim$}}}}

\nc{\etal}{{\it et al.}}
\nc{\Lsp}{\;\;\;\;\;\;\;\;\;\;}  \nc{\LLLsp}{\lspace \lspace}
\nc{\lsp}{\;\;\;\;\;\;}
\nc{\spac}{\;\;\;}
\nc{\noi}{\noindent}
\nc{\beq}{\begin{equation}}   \nc{\eeq}{\end{equation}}
\nc{\bea}{\begin{eqnarray}}   \nc{\eea}{\end{eqnarray}}
\nc{\baa}{\begin{array}}      \nc{\eaa}{\end{array}}
\nc{\bit}{\begin{itemize}}    \nc{\eit}{\end{itemize}}
\nc{\ben}{\begin{enumerate}}  \nc{\een}{\end{enumerate}}
\nc{\bce}{\begin{center}}     \nc{\ece}{\end{center}}
\nc{\red}{\textcolor{red}}

\def\sq2{\sqrt{2}}

\def\ph{\varphi}

\def\m4{m^4(\ph)}
\def\mn2{M_n^2}

\def\v5{V^{(5)}}

\begin{document}

\vspace{.5cm}

\title{
%Bulk Higgs leaking out of the brane\\
%Delocalizing the Higgs of warped extra dimensions\\
%   An odd Higgs in lower-scale warped extra dimensions\\
% An unusual Higgs  in lower-scale  warped extra  dimensional models\\
%   A wavy Higgs in lower-scale warped extra dimensions\\
%Higgs and Dirichlet and a warped extra dimension\\
A nearly Dirichlet Higgs for lower-scale warped extra dimensions
 }
\author[1]{Mariana Frank}
\ead{mariana.frank@concordia.ca}
\author[2]{Nima Pourtolami}
\ead{nima.pourtolami@gmail.com}
\author[3]{Manuel Toharia}
\ead{mtoharia@dawsoncollege.qc.ca}

\affiliation[1]{organization={Department of Physics,
  Concordia University}, addressline={7141 Sherbrooke St. West},
  city={Montreal}, postcode={Quebec H4B 1R6}, country={Canada}}

\affiliation[2]{organization={National Bank of Canada}, addressline={1155 Metcalfe St.},
  city={Montreal}, postcode={Quebec H3B 4S9}, country={Canada}}

\affiliation[3]{organization={Department of Physics, Dawson College},addressline={3040 Sherbrooke
  St W}, city={Montreal}, postcode={Qc, H3Z 1A4}, country={Canada}}

\date{\today}

%\maketitle

\begin{abstract}
We consider a minimal extension of the Standard Model in warped
extra dimensions, with fields propagating in the bulk including a bulk
SM-like Higgs doublet. 
We show that the Higgs can acquire a non-trivial oscillatory VEV,
strongly localized towards the TeV brane, but such that its value at
that brane could be highly suppressed due to its oscillatory behaviour. Within
the minimal Randall-Sundrum metric background, this oscillatory  VEV
 can alleviate the bounds coming from oblique precision
electroweak parameters, such that the KK gluon mass can be around $3$
TeV (instead of $\sim 8$ TeV for the usual 
non-oscillatory bulk Higgs). We also discuss the stability of the
configuration as well as the naturalness of the model parameters. 

\end{abstract}

\maketitle

%%%%%%%%%%%%%%%%%%%%%%%%%%%%
\section{ Introduction}
\label{sec: introduction}
%%%%%%%%%%%%%%%%%%%%%%%%%%%%
The original motivation for warped extra dimensional models was to
solve the weak-Planck scale hierarchy by allowing gravity to propagate
in the bulk of the extra dimension
\cite{Randall:1999ee,Randall:1999vf}, stabilized by a Goldberger-Wise
type mechanism
\cite{Goldberger:1999uk,Goldberger:1999un,Csaki:2000fc,Csaki:1999mp,DeWolfe:1999cp,Arkani-Hamed:1998cxo}.
Furthermore, by allowing the SM fermion fields to propagate into the bulk, it was
found that the localization of fields along the extra dimension could
provide an explanation for the observed masses and flavor mixing among
quarks and leptons  \cite{Davoudiasl:1999tf, Pomarol:1999ad,
  Grossman:1999ra,Chang:1999nh, Gherghetta:2000qt,
  Davoudiasl:2000wi}. In these scenarios, while the electroweak
symmetry breaking can still proceed through the standard Higgs
mechanism, the AdS/CFT correspondence can also be used so that the
Higgs appears as a composite pseudo-Goldstone boson of the strongly
coupled theory \cite{Contino:2003ve,Csaki:2008zd}. In these models the
Higgs boson field must be localized near the TeV boundary of the extra
dimension in order to solve the hierarchy problem. However, it is also
possible for the Higgs to leak into the bulk (bulk Higgs
scenario). The benefit of this scenario is to alleviate some of the flavour bounds
and precision electroweak tests plaguing the brane Higgs models
 \cite{Carena:2004zn, Huber:2003tu,
  Agashe:2004cp, Agashe:2006at, 
  Csaki:2008qq}. Generally, in these models, the bounds from the
precision flavour and electroweak processes push the mass scale of
new particles to relatively large scales, no lower than $\sim 8$ TeV, which is not accessible to experiments
yet. Two different solutions have been proposed to satisfy these
bounds, while allowing for light enough new physics within the reach
of the LHC: one is to extend the gauge group  to include custodial
symmetry \cite{Agashe:2003zs,Agashe:2006at}, the other is to modify
the warping of the space-time metric away from the pure AdS spacetime
\cite{Falkowski:2008fz,Batell:2008me,Cabrer:2011fb,Carmona:2011ib,MertAybat:2009mk}. 

In this letter, we point out that the parameter space of the usual bulk
Higgs, with no gauge group extensions or modified metric backgrounds,
has not been fully explored. We study here a remaining 
region of parameter space and show that the bounds from precision
electroweak tests are much less constraining in that region. The
metric background is unchanged and the gauge groups are the usual 
SM groups, however the nontrivial Higgs VEV that we consider can have a
substantial suppression on the TeV boundary.

%%%%%%%%%%%%%%%%
\section{The odd Higgs}
\label{sec:model}
%%%%%%%%%%%%%%%%%%%%%%%%%%%%%%%%%%%%%%%%%%%%%%%%%%%
We consider a scenario with one extra space dimension and assume a
(properly stabilized) static spacetime background as 
\bea
ds^2 = e^{-2\sigma(y)}\eta_{\mu\nu} dx^\mu dx^\nu - dy^2\, ,
\label{RS}
\eea
where $\sigma(y)$ is a warp factor responsible for the exponential
suppression of the mass scales from the UV brane, down to the IR
brane, located at the two boundaries of the extra coordinate, $y=0$
and $y=y_{1}$, respectively  \cite{Randall:1999ee,Randall:1999vf}.
We  assume that the dynamics responsible for stabilizing the setup ({\it i.e.} giving
mass to the radion) do not back-react strongly on the background
metric so that we can use the simple Randall-Sundrum (RS) metric with the warp
factor given by $\sigma(y)= k y$. The parameter $k$ is assumed to be
of the order of the fundamental scale $M_{\rm Planck}$,  so that $k y_1 \simeq
34$   and the  hierarchy between the Planck and the electro-weak scale  is
achieved  naturally, and we will refer to the warped-down scale
$\Lambda_{KK}=ke^{-ky_1}$ as the Kaluza Klein (KK) scale.  

The matter content of this scenario corresponds to a minimal 5D extension of the
SM gauge group  and with all fields propagating in the bulk
\cite{Davoudiasl:1999tf,Grossman:1999ra,Pomarol:1999ad}, such that
fermion field localizations along the extra dimension can address the flavor puzzle of the SM \cite{Agashe:2004cp}.

Electroweak symmetry breaking (EWSB) is induced by a single 5D bulk Higgs
doublet ($H$) appearing in the Lagrangian density as
\bea
&&\hspace{-1.5cm}{\cal L}_5 \supset \sqrt{g}\ \left(-\frac{1}{4g_5^2}W_{MN}^2-\frac{1}{{g_5'}^2}B_{MN}^2 - |D^M H|^2 - V(H) \right)  \non\\
&&-\sqrt{g}\ \Big (\delta(y)  \lambda_0(H) +  \delta(y-y_1)\lambda_1(H) \Big) ,  
\label{eq:5Daction}
\eea
where the capital indices run through the $5$ spacetime
directions, $M = (\mu, 5)$, while the Greek indices, $\mu, \, \nu, ...$,  denote the usual 4D
dimensions \cite{delAguila:2006atw}.  We first consider the simple case of a quadratic bulk Higgs
potential with the addition of two brane localized potentials,
such that the one located at $y=y_1$ triggers EWSB. We define  the parameters in Eq. (\ref{eq:5Daction}) as 
\bea
 V(H)&=& \frac{1}{2}\mu_B^2 H^2 \,,\\
\lambda_0(H)&=&\frac{1}{2}M_0H^2\,,\\
 \lambda_1(H)&=&\frac{1}{2}M_1 H^2+\frac{1}{2}\gamma_1 H^4\,.
\eea

The 5D Higgs doublet can be expanded around a nontrivial VEV profile
$v_{odd}(y)$ in a similar way as in the SM
\bea
\label{Hexpansion}
H=\frac{1}{\sqrt{2}} e^{i g_5\Pi  }\left(\begin{matrix} 0\\ v(y) +h(x,y) \end{matrix}
\right) \, .
\eea
Within the simple RS metric,  the static non-trivial
Higgs VEV profile, $v_{odd}(y)$, has to satisfy the bulk equation
\bea
 v'' -4 k v' - \mu^2_B  v & =& 0,
 \label{yvev}
 \eea
and to obey the boundary conditions 
 \bea
&v'(0) &=  M_0\ v(0)\,, \\
&v'(y_1) &=  -M_1 v(y_1) -2 \gamma_1 v(y_1)^3 \,.
\label{vbc}
\eea
There are two different types of nontrivial solutions\footnote{The
  trivial solution $v=0$ is always possible. This  corresponds to the electroweak
  symmetric background and, if it is unstable, it will trigger  EWSB.} to Eq. (\ref{yvev}),
depending on the size of the bulk mass parameter.\footnote{Note that
  the sign of the real parameter $\mu^2_B$ can in principle be either positive or negative.}
 We distinguish two cases:
\bit
\item If $(\mu_B^2 \geq -4 k^2$) then the nontrivial solution is
  \bea
  \hspace{-.8cm}v_{usual}(y)=v_0  \left(e^{a k y} - \frac{M_0/k-a}{M_0/k-4+a} e^{(4-a) ky} \right)
  \label{vusual}
  \eea
  where we have introduced the real parameter $a$ as $a k=2k+\sqrt{\mu_B^2
    + 4k^2}$   and we have imposed the boundary
  condition at $y=0$. The boundary condition at $y=y_1$ yields
  and equation for the amplitude $v_0$ in terms of the parameters $k$, $M_0$, $M_1$,
  $\gamma_1$ and $\mu_B^2$. This solution is the usual nontrivial VEV
  solution used in the literature when considering a bulk Higgs
  mechanism.
\item If $(\mu_B^2 \leq -4 k^2$) then the nontrivial solution is
  \bea
  \label{v_odd}
 \hspace{-.8cm} v_{odd}(y)=v_0 e^{2 k y} \left(\sin{(b k y)} + \frac{b}{(M_0/k-2)} \cos{(b ky )}\right)
  \eea
  where the real parameter $b$ is defined as
  $b k =\sqrt{-\mu_B^2 - 4k^2}$ and we have imposed the boundary
  condition at $y=0$. The second boundary condition at $y=y_1$ will constrain
  the amplitude $v_0$ in terms of the parameters $k$, $M_0$, $M_1$,
  $\gamma_1$ and $\mu_B^2$.
This solution can yield a Dirichlet-like and more delocalized (but
still solving the hierarchy problem) bulk Higgs, 
  which we refer to subsequently as the ``odd'' Higgs.  
\eit
  Even though the last solution does not necessarily have {\it
      odd} parity under orbifold reflection ({\it i.e.} exact Dirichlet
  boundary conditions), it contains that  possibility. The most
  interesting region in the parameter space  of this novel VEV profile
  is where the first term dominates and hence, the solution is almost
  odd (or Dirichlet-like). 
As we will see shortly, stability related bounds on the free
parameter $b$ (fixed by the  bulk Higgs mass, $\mu_B^2$), are such
that $b \leq   \pi/(ky_1)$. It is therefore generally expected that
$b\ll (M_0/k-2)$,  and hence, Eq. (\ref{v_odd}) reduces to $v_{odd}(y) 
\simeq v_{odd}^{(--)}(y)$, where we refer to the
pure $sine$ solution as the $(--)$ odd Higgs profile
$v_{odd}^{(--)}(y)=v_0 e^{2ky}  \sin{(bky)}$.  In the case where $b \gg (M_0/k-2)$ (possible
whenever $M_0/k$ is really close to $2$), the profile
becomes $v_{odd}(y)  \simeq v_{odd}^{(+-)}(y)$, where now we refer to
the pure $cosine$ solution as the $(+-)$ odd  Higgs solution, $v_{odd}^{(+-)}(y)=\tilde{v}_0 e^{2ky}
\cos{(bky)}$, where $\tilde{v}_0$ is the proportionality constant of
this solution.

%% {\color{blue} $v_{odd}(y)=v_0 e^{2 k y} \frac{b}{(M_0/k-2)} \cos(b
%%   ky )$. Am I missing something? where did the $ \frac{b}{(M_0/k-2)}$
%%   go? do we redefine $v_0$?} 

We  must mention here that $\mu^2_B<-4k^2$
violates the Breitenlohner-Freedman bound for the full AdS space
\cite{Breitenlohner:1982jf}.  
However, for a $slice$ of AdS, the boundary effects
slightly modify the stability threshold  \cite{Toharia:2008ug}, 
and therefore, it is possible for $\mu^2_B$ to be slightly below
$-4k^2$. Stability considerations still impose 
strict constraints on $\mu_B^2$ and in particular, on the VEV which should not have any
nodes within the interval
\cite{Grzadkowski:2004mg,Toharia:2007xe,Toharia:2007xf,Toharia:2010ex}.
This implies that in order to obtain stable, oscillatory VEV solutions we must have\footnote{Strictly speaking,
  this is the condition for the existence of stable, nontrivial solutions with
  Dirichlet boundary conditions on both branes.}:
 $-4k^2-\pi^2/y_1^2\leq\mu^2_B\leq - 4k^2$.
 This condition is the origin of the bound, $b \leq \pi/(ky_1)$  mentioned earlier, and
since $(ky_1) \simeq 34$, 
we require: $0\leq b \lesssim 0.1$.  We will return to stability
issues in Section \ref{sec:stability}. 

 Next, we consider the physical Higgs perturbations
$h(x,y)=\sum_n h^n_x(x) h_n(y)$ around the nontrivial  VEV. The
profiles $h_n(y)$ of the Higgs (KK) modes  satisfy the
equation
\bea
h''_n - 4k h'_n - \mu_B^2 h_n + m_n^2 e^{2 ky}h_n=0,
\label{h_eq}
\eea
with boundary conditions
\bea
& h_n'(0) &=  M_0\ h_n(0)\,, \\
& h_n' (y_1) &=  - (M_1 -6 \gamma_1 v(y_1)^2) h_n(y_1).
\label{h_bc}
\eea
It is apparent that with $m_n=0$, the bulk Higgs profile and the
background VEV satisfy identical equations of motion but with different boundary 
conditions. (Although for our choice of brane potentials, the
boundary conditions at $y=0$  are the same). 

Assume that the mass $m_0^2$ of the lowest mode (which will be
identified as the SM Higgs) is small compared to the KK scale, so that
the problem can be treated perturbatively. We have
\bea
h_0(y)=\alpha\ v(y)\ (1+{\cal O}(m_0^2/\Lambda_{KK}^2))\,,
\label{handv}
\eea
where the KK scale is as before $\Lambda_{KK}=k e^{-k y_1} \sim
{\cal O}$(TeV), and with $\alpha$ a proportionality factor to be fixed
by the canonical normalization of the Higgs mode. Then, the Higgs mass can be written as (see
for example \cite{Cabrer:2011fb})
\bea
\hspace{-.7cm}m_0^2=-\frac{2 v^2(y_1) e^{-4
    ky_1}\left(M_1+\frac{v'(y_1)}{v(y_1)}\right)}{\int_0^{y_1}dye^{-2k
    y} v^2(y)} \left(1+{\cal
  O}\left(\frac{m_0^2}{\Lambda_{KK}^2}\right)\right) \, , 
\eea 
which is valid for perturbations around any nontrivial VEV.
In particular, perturbing around $v_{usual}(y)$ one finds
\bea
(m_{0}^2)_{usual} \simeq - 8 (a-1)\ \left(\frac{M_1}{2k}+ \frac{a}{2}\right)\ \Lambda_{KK}^2,
\label{mbulk}
\eea
with $a\geq2$. This result shows the % $(10^{-2})-(10^{-3})$
amount of tuning needed in the usual bulk Higgs setup, since, 
in order to satisfy Eq. (\ref{mbulk})
%depending on the KK scale, $(m_{0}^2)_{usual}/\Lambda_{KK}^2$ is 
%around  $0.01$ to $0.001$,
%physical Higgs mass $squared$ must be about two orders of magnitude
%smaller than the KK scale $squared$. 
we must first require
$M_1$ to be negative (for the Higgs
mass to be positive) and second, an ${\cal O}(1\%-0.1\% ) $ cancellation between $M_1/(2k)$
and $a/2$, which are, in principle, independent parameters, originating from
the brane and  bulk Higgs potentials respectively. 

Moving on to the case of the ``odd'' Higgs, we take\footnote{Here, for simplicity, we assumed $b\ll (M_0/k-2)$. This will not
  change any qualitative features as we will see later.}
  $v_{odd}^{(--)}$%=v_0 e^{2ky}\sin{(bky)}$,
  and obtain
  \bea
(m_{0}^2)_{odd} \simeq - 8 (1+b^2) F(b)\  \Lambda_{KK}^2,
  \label{modd}
    \eea
where we have defined
  \bea
  F(b)=\frac{\sin^2{(bky_1)} (2+\frac{M_1}{k}+b
    \cot{(bky_1)})}{1+b^2- \cos{(2bky_1)}-b \sin{(2bky_1)}}.
  \eea
This expression is only valid for small $m_0^2$, and in general it
is possible to fix $M_1$ such that $F(b)$ is suppressed.

We can also see that suppression of $\sin{(bky_1)}$ at the boundary 
(i.e, $b ky_1\sim \pi$), will lead to a suppression of the
Higgs mass and we find 
\bea
(m_{0}^2)_{odd} \simeq  8  \frac{\sin{(bky_1)}}{b}  \Lambda_{KK}^2.
\label{mhapprox}
\eea
%where we have assumed $\sin{(bky_1)} \ll 1 $ (and $\cos{(bky_1)}
%\sim-1$).
Moreover, the sign of  $\sin{(bky_1)}$  indicates whether  the
Higgs mass is tachyonic or not. In particular, for 
 $bky_1=(\pi+\epsilon)$, with $\epsilon$ small,
 the Higgs mass becomes tachyonic. 
% while  if $bky_1 =(\pi+\epsilon)$  the Higgs
%     mass squared is negative. 
     When $bky_1=\pi$, the Higgs will be
     exactly massless. This result agrees with the stability criterion
     introduced earlier, since $bky_1=\pi+\epsilon$ 
 means that the  corresponding VEV would have a node within the
interval.
%% Also, since $b \lesssim 10^{-1}$, we see from Eq. (\ref{mhapprox})
%% that the Higgs VEV suppression at the TeV brane must be of the order
%% of  $\sin{(bky_1)} \lesssim (10^{-3}-10^{-4})$ in order to obtain a
%% light enough Higgs mass.

 %% Thus, in order to  obtain a light Higgs mass, this is  the level of
 %% fine tuning required for the ``odd'' Higgs, which has moved from a
 %% tuning between $M_1/k$ and $a$, into a tuning between $ky_1$ 
 %% and $b$ (or equivalently a tuning between $1/y_1^2$ and $\mu_B^2$).
 %% Before further discuss stability and naturalness  
 %% in Section \ref{sec:stability},

Next, we analyse  how this odd VEV improves the precision
electroweak bounds on the model parameters.
  
%%%%%%%%%%%%%%%%%%%%%%%%%%%%%%%%%%%%
\section{Precision Electroweak tests}
\label{sec:ST}
%%%%%%%%%%%%%%%%%%%%%%%%

We now consider  the effects of  integrating out the gauge KK modes on
the precision electroweak parameters by calculating  the corrections
to the $T$ and $S$ parameters \cite{Carena:2004zn}. These can be easily computed with  a set of surprisingly compact
and simple integrals (see \cite{Cabrer:2011fb} for details), given by
\bea
\hspace{-1.3cm}&& \alpha T= s_W^2 m_Z^2 y_1  \int_0^{y_1}dy e^{2ky}
(1-\Omega_h(y))^2             \\ 
\hspace{-1.3cm}&& \alpha S= 8 s_W^2c_W^2 m_Z^2 y_1 \int_0^{y_1}dy
e^{2ky}\left(1-\frac{y}{y_1}\right) (1-\Omega_h(y)) \ \
\eea
with $s(c)_W$ being $\sin \theta_W (\cos \theta_W)$, and the function $\Omega_h(y)$ defined in terms of the
light SM-like Higgs profile, $h_0(y)$, as: %$h_0$ as
\bea
\Omega_h(y)=\frac{\int_0^ydy' e^{-2 ky'}h_0^2(y')}{\int_0^{y_1} dy'
  e^{-2 ky'}h_0^2(y')}\, . 
\eea
 Note that because the mass of $h_0$ is small compared to the KK scale, its wave function is
proportional to the nontrivial  VEV $v_{odd}(y)$, {\it i.e.} $h_0(y)\propto
v_{odd}(y)$. We consider the two different VEVs discussed in the previous
section, namely the usual bulk Higgs VEV and the odd Higgs VEV, 
\bea
v_{usual}(y)&\simeq& v_0 e^{a k y}
\label{eq:vusual}
\\
 v_{odd}(y)& \simeq& v_0 e^{2ky}\sin{(bky)}
 \label{eq:vodd}
\eea
where, for simplicity, we have only kept the leading term in each case.
%order to simplify the analytical expressions, which we present here. 
Full formulations are easily obtained, and used only for
later numerical computations, since they do not affect our discussion here.
%improve the bounds obtained here. 

One can find expressions for the $T$ and $S$ parameters  in each
case, by evaluating the integrals analytically. We  ignore
warped down terms, and also assume that $ v_{odd}(y)$ is highly
suppressed near the TeV brane.
%{\it i.e.} $\sin{(bky_1)}\sim 10^{-4}$, and
%therefore neglect  higher  order terms in $\sin{(bky_1)}$  and  in
%$b^2\sim 10^{-2}$. 
We find 
\bea
\alpha T_{usual}&=&   s_W^2 \frac{m_Z^2}{\Lambda_{KK}^2}  (ky_1)   \frac{(a-1)^2}{a(2a-1)}\,,\\
\alpha S_{usual}&=&8 s_W^2c_W^2 \frac{m_Z^2}{\Lambda_{KK}^2} \frac{a^2-1}{4a^2}\,, 
\eea
and
\bea
\label{Todd}
\alpha T_{odd}&=&  s_W^2 \frac{m_Z^2}{\Lambda_{KK}^2}  (k y_1)
\frac{17}{648}\ \left(1+{\cal O}(b^2,\epsilon_b)\right) \,, \\ 
\label{Sodd}
\alpha S_{odd}&=&8 s_W^2c_W^2 \frac{m_Z^2}{\Lambda_{KK}^2}
\frac{5}{64} \ \left(1+{\cal O}(b^2,\epsilon_b)\right) \,,  
\eea
where $\epsilon_b$ is defined as $b=\pi/(ky_1) - \epsilon_b$, and assumed to be small, 
so that the Higgs profile is almost Dirichlet on the $y=y_1$ boundary. 
Because $\mu^2_B/k^2 = -4-b^2$, a
very small  value for $\epsilon_b$  means that the bulk Higgs mass parameter
$\mu^2_B$ is very close to the stability limit of $-4k^2-\pi^2/(y_1)^2$.% (in units of units of $k^2$).

\begin{figure}[t]
\center
\begin{center}
  \includegraphics[height=6.8cm,width=8.5cm]{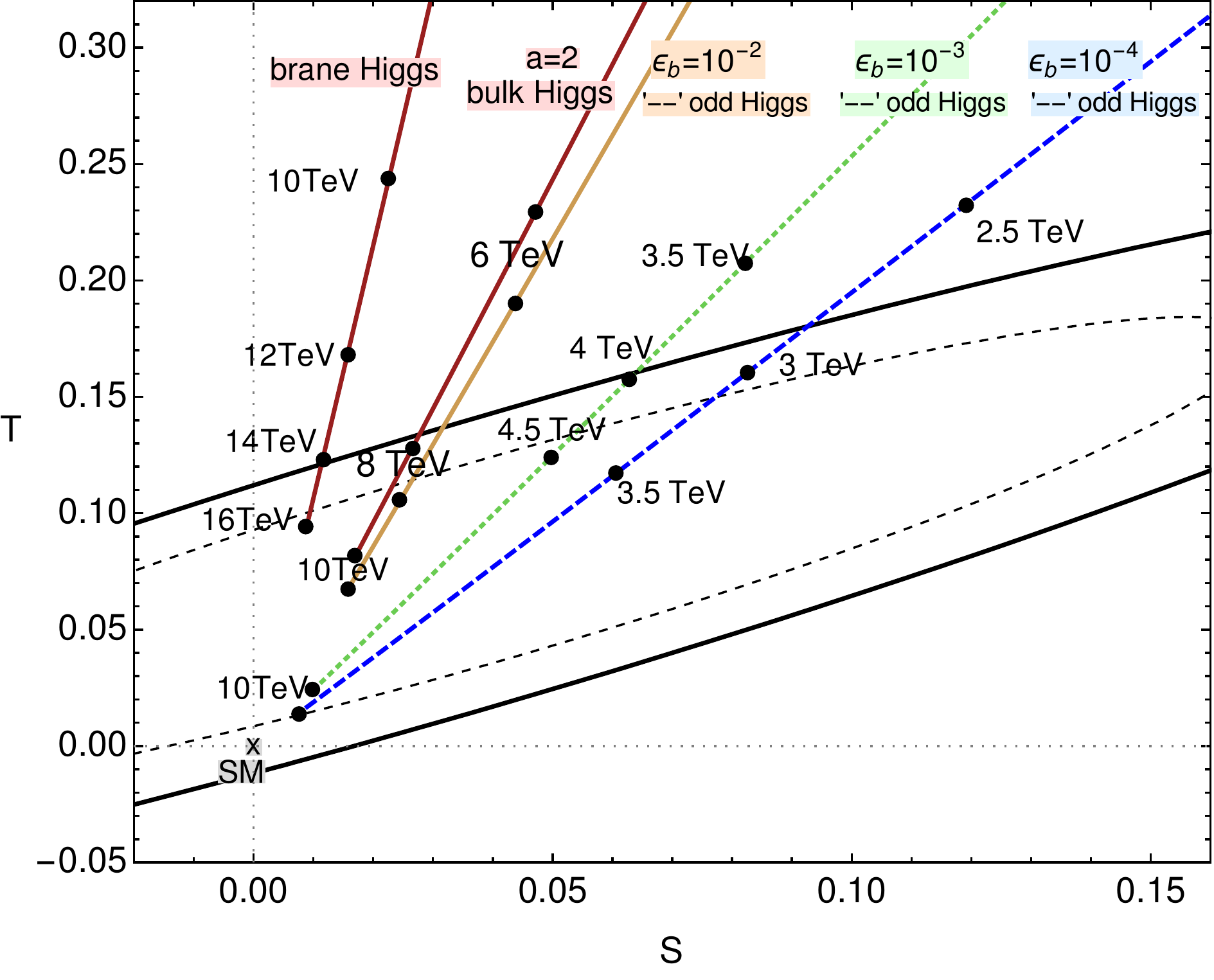}\ \ 
\end{center}
\vspace{-.2cm}
\caption{
%% \color{green} Inner (outer) {\color{blue} please check, I changed
%% the order} ellipse for the allowed region in the $S-T$ 
  %% parameter plane at the 95\% CL (68\% CL) as  dashed (solid) black curves \cite{Haller:2022eyb}, 
  %% %  as enclosed in the dashed (solid) black curves,
  %% We could jusst write (and it is shorter)}\\
  Allowed region in the $S-T$ parameter plane at the 95\%  and 68\%  CL  \cite{Haller:2022eyb}, 
  as well as the $S$ and $T$
  predictions (from tree-level KK gauge boson exchange) for a
  bulk Higgs scenario in the three regimes of brane Higgs, bulk Higgs
  and odd Higgs.  Each curve is obtained by varying the KK scale which
  we parametrize using the physical mass of the lightest KK gauge
  boson $(M_{KK})_1$.  }
\label{fig:ST}
\vspace{-.2cm}
\end{figure}

 We  focus first on the $T$ parameter as,    due to the volume factor enhancement of
$(ky_1)\sim 34$,  it is the most constraining of the oblique parameters. The usual bulk 
Higgs result depends on the parameter $a$, and is such that the least
constrained result is obtained for $a=2$ ({\it i.e.} a relatively
delocalized Higgs VEV), where the numerical factor from the integral becomes $1/6$. When $a$ is
very large (corresponding to a brane localized Higgs VEV), the factor becomes $1/2$.  %(three times larger). 
Therefore the bound on the KK scale coming from the
$T$ parameter is $\sqrt{3}$ times smaller for a delocalized bulk Higgs 
as compared to a brane Higgs. The same integral factor in the odd Higgs case is
$17/648$, meaning that the
bound on the KK scale is $\sqrt{648/17\times 6}\sim2.5 $ times better ({\it i.e} weaker) for the
odd Higgs case compared to the most delocalized usual Higgs case $(a=2)$.

For the less constraining $S$ parameter, again, the odd Higgs  
corrections are more suppressed. 
%in comparison to the to the $a=2$ bulk Higgs. 
However, here the $S$ parameter correction is
larger relative to the $T$ parameter correction. Indeed  comparing
the two cases again
\bea
\hspace{-.5cm}\frac{T_{usual}}{S_{usual}}=\frac{ky_1}{8c_W^2}  \frac{4 a^2
  (a-1)^2}{(a^2-1)a(2a-1)} \sim 0.89 \frac{ky_1}{8c_W^2}\ \ (a=2) \,,
\eea
and
\bea
\frac{T_{odd}}{S_{odd}}=\frac{ky_1}{8c_W^2} \frac{136}{405}\  \sim\  0. 34\frac{ky_1}{8c_W^2} \, .
\eea
Thus  for the
odd Higgs case the $S$ parameter is relatively more important.  The effect is to push the
overall electroweak corrections towards a more favourable
direction in the $S-T$ plane, allowing for lower KK scales
than those allowed by only considering the $T$ parameter.\footnote{Note
  that this effect is already present for the usual bulk Higgs, since the
  ratio $T/S$ is about two times larger for a brane Higgs ($a$ large)
  compared to a delocalized Higgs ($a=2$). Therefore the bounds for
  $a=0$  shift towards a more favorable direction in the $S-T$ plane and are thus improved
  further compared to the most constraining brane  Higgs.}
This is shown in Figure \ref{fig:ST}, where
%we show the outer (inner) ellipse, as enclosed in the dashed (solid) black curves, for the allowed region in the S-T
%  parameter plane at the 95\% CL (68\% CL) from \cite{Haller:2022eyb}. 
 the predicted values for $S$ and $T $ arising
from integrating out the KK electroweak bosons are traced for different values of  
the bulk Higgs parameter, $\mu_B^2$, by varying the KK scale of the
model, parametrized as the lightest KK gauge boson mass.
As shown earlier, it is useful to parametrize this scale using the $a$ and $b$ parameters. When $a$ is large
($\mu_B^2$ large and positive) the Higgs sector  corresponds to the brane Higgs
limit and the bounds from the oblique parameters are the strongest. In this case
 the lightest KK gluon cannot be lighter than $\sim 14$ TeV at
$95\%$ CL. As the Higgs leaks out of the brane, by reducing $a$ (or
$\mu^2_B$), the bounds improve. At $a=2$ ($\mu^2_B=-4k^2$) 
the usual limit of bulk Higgs delocalization is obtained and the bounds are such
that the KK gluon mass should not be lower than $\sim 8$ TeV. As
discussed earlier, the bulk Higgs mass can be lowered  beyond $-4k^2$
as long it respects the stability bound $-4k^2-\pi^2/y_1^2 \leq \mu_B^2$. 
%Because $(ky_1)\sim 34$, the extra parameter space is restricted, so we
%parametrize it using the small parameter $\epsilon_b$, defined as the
%relative mass difference between  $\mu_B^2$ and the stability limit. 
The parameters $S$ and $T$ are evaluated numerically and show a  striking
improvement in their bounds, as the parameter $\epsilon_b$ is reduced,
({\it i.e.}, as the Higgs is more and more Dirichlet-like). There are, however, limiting
values for each ({\it cf.} Eqs. (\ref{Todd}) and (\ref{Sodd}), so that once
$\epsilon_b$ is small enough,  no further improvement can be obtained. Within
the odd Higgs regime,  the lowest mass of the KK gluon,
consistent with the $S-T$ bounds is about $2.8$ TeV. This represents the
least constraining (calculable) limit on RS models without custodial
symmetry or large metric back-reaction \cite{Cabrer:2011fb,Cabrer:2009we,
  Cabrer:2010si,Quiros:2013yaa,Carmona:2011ib,Dillon:2014zea}. 

\begin{figure}[h]
\center
\begin{center}
  \includegraphics[height=6.8cm,width=8.5cm]{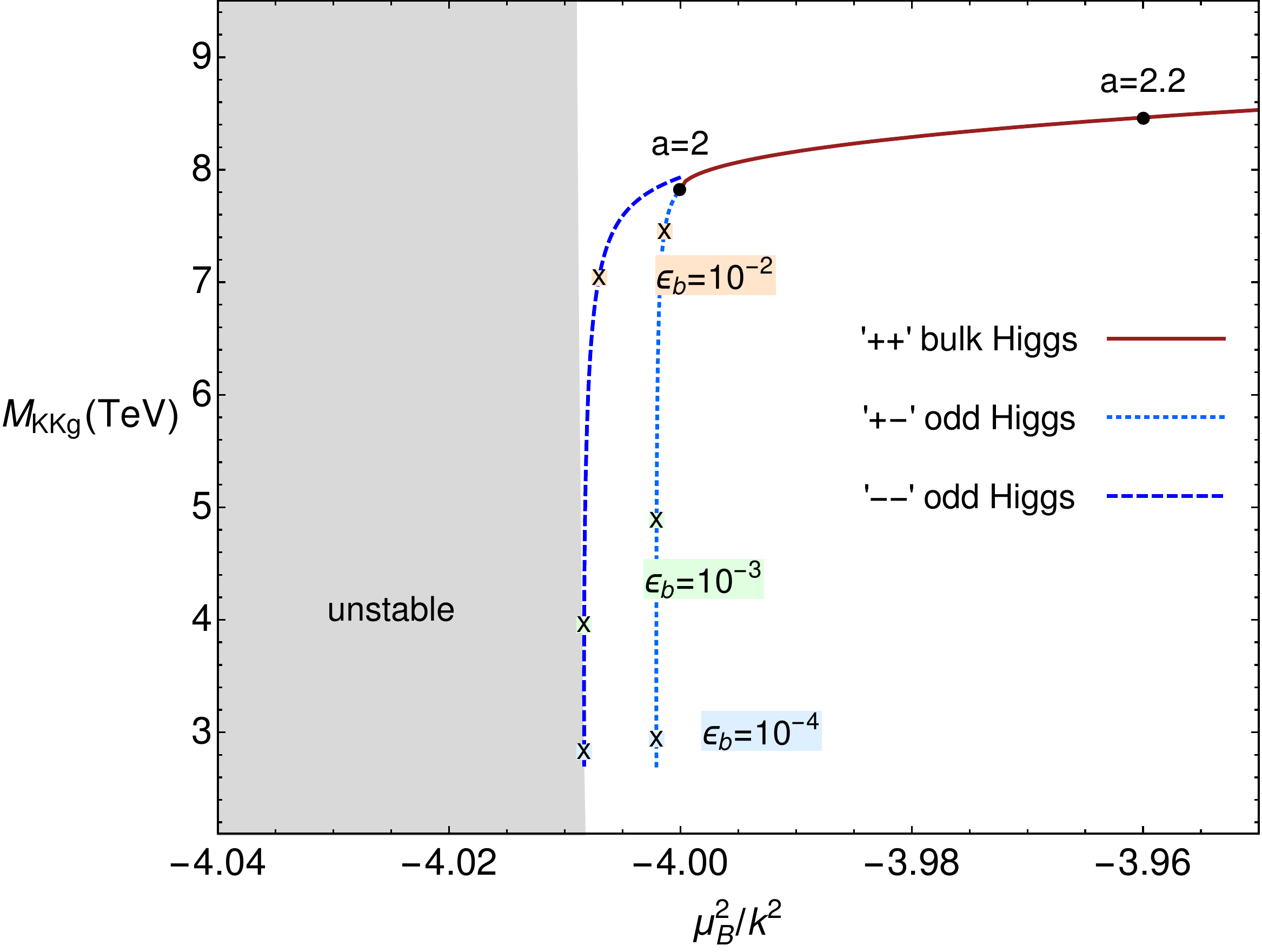}\ \ 
\end{center}
\vspace{-.2cm}
\caption{95$\%$ CL lower bound from $S$ and $T$ parameters constraints on the mass of the first KK gauge boson
  as a function of the (negative) bulk Higgs mass coefficient,
  $\mu_B^2$. The grey region at the left side represents the region (in $\mu_B^2/k^2$) 
  where the bulk Higgs mass parameter is smaller than the lower bound
  set by stability requirements and is thus ruled out. 
  % .  When $-4k^2>\mu_B^2\geq-4k^2-\left(\frac{\pi}{ky_1}\right)^2$ the scenario
 % is in the odd Higgs regime with
 The parameter $\epsilon_b$ denotes the mass difference (in units of
 $k^2$) between $\mu_B^2$ and the stability limit. When 
$\mu_B^2\geq -4k^2$, the scenario enters the usual bulk Higgs regime
and is commonly parametrized by the Higgs wave function coefficient $a\geq2$.}
\label{fig:mkkvsmub2}
\vspace{-.2cm}
\end{figure}

In Figure \ref{fig:mkkvsmub2}, we consider again  the tree level
corrections to $S$ and $T$ from KK gauge bosons, but in this case we show
the lowest possible KK gluon mass as a function of the bulk Higgs mass,
$\mu^2_B$, near the %threshold region of
$\mu_B^2=-4k^2$ region. As expected,
the bounds  improve slightly as the Higgs mass approaches $-4k^2$ from
above. Once the mass decreases beyond that point, the bounds
 improve dramatically  until reaching the stability limit. Note that, as mentioned earlier, the
stability limit differs depending on the UV boundary conditions
on the odd Higgs VEV. When the UV brane mass parameter, $M_0$, is large
enough, the Higgs acquires the $(--)$ nontrivial VEV, while 
%$v^{(--)}_{odd}(y)=e^{2ky}\sin{bky}$. 
 if there is a strong
cancelation $(M_0/k-2)\ll1$ the VEV becomes the $(+-)$ VEV.
%$v^{(+-)}_{odd}(y)=e^{2ky}\cos{bky}$.
In this case, the condition for the VEV to have no nodes within the interval is
$b\leq \pi/(2ky_1)$, which yields a slightly different bound for
  $\mu_B^2$. This appears in the figure as a vertical line for the
  $(+-)$ case, appearing at a different value of $\mu^2_B$ than the vertical asymptote
  for the $(--)$ case. Note that the best bound on the KK scale is the
  same in both situations, and when the odd Higgs VEV has an
  expression containing both $\sin{(bky)}$ and $\cos{(bky)}$, the stability limit of $\mu^2_B$
  will lie between the limits of the $(--)$ and the $(+-)$ case. In the
  figure we show the region set
  by the $(--)$ solution as the grey region marked as unstable. 
\begin{figure}[h]
\center
\begin{center}
  \includegraphics[height=6.8cm,width=8.5cm]{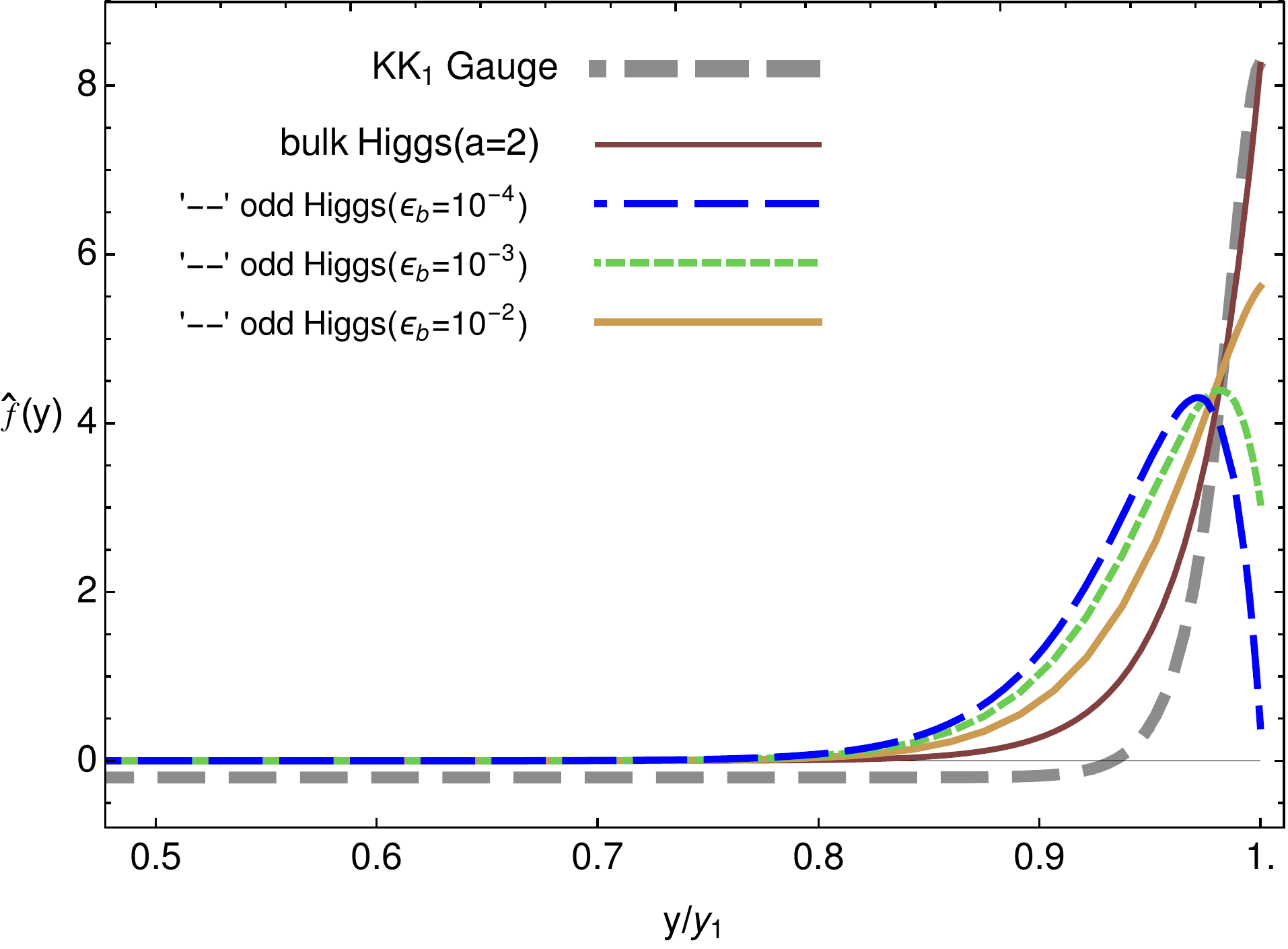}\ \ 
\end{center}
\vspace{-.2cm}
\caption{Profiles of the ``physical'' wave-functions, $\hat{f}(y)=e^{-ky} v_{odd}(y)$, of
  the lightest Higgs mode in the bulk Higgs regime with $a=2$
  (solid, dark brown) and in the odd Higgs regime with $\epsilon_b=10^{-4}$
  (dashed, blue), $\epsilon_b=10^{-3}$ (dotted, light green) and
  $\epsilon_b=10^{-2}$ (light orange). Also shown is the wave-function
  of the first heavy KK gauge mode (grey, thick dashed). The overlap between
  Higgs and KK gauge bosons is clearly suppressed as the Higgs wave function becomes
  diminished near the brane. This leads to a suppressed contribution
  to the $T$ parameter.}
\label{fig:wavefunctions}
\vspace{-.2cm}
\end{figure}

Finally,  in Figure \ref{fig:wavefunctions}, we show the
Higgs profiles (and therefore VEVs) for different values of the bulk Higgs mass, $\mu_B^2$.
%in different regimes defined
%by different values of the bulk Higgs mass $\mu_B^2$ ({\it i.e.} 
%for different values of
%either the $a$ parameter or the $\epsilon_b$ parameter. 
%Note that the wave-functions are normalized, and include the metric warp factor $e^{-ky}$.
%Note that we
%are plotting the normalized wave function multiplied by the metric
%factor $e^{-ky}$ accompanying each Higgs field entering the kinetic term in the Lagrangian. 
The profile of the lightest KK gauge mode (same in all
regimes) is also plotted for comparison. 
We observe that, as the Higgs mode
becomes more and more Dirichlet-like, it leaks out of the boundary ({\it cf.} \cite{Quiros:2013yaa,Carmona:2011ib}),
%a suppression of this type of the Higgs physical wave function 
which will lead to weaker couplings with
the KK gauge bosons and thus suppressed
contributions to the $S$ and $T$ parameters.

%Before  discussing existence and stability of the odd
%Higgs VEV we quickly overview the Yukawa couplings.
%%%%%%%%%%%%%%%%%%%%%%%%%%%%%%%%%%%%%%%%%%%%
\section{Fermion masses and Yukawa couplings}
\label{sec:Yukawa}

Consider the following 5D up-quark sector Yukawa Lagrangian density
\bea
&&\hspace{-1cm} {\cal L}_{Y}= \sqrt{g} \left(
\frac{Y_u^{bulk}}{\sqrt{k}}  HQU\right)
%+\frac{Y_d^{bulk}}{\sqrt{k}} HQD %\right)
+ h.c. \,,
\label{5DYukawas}
\eea
where the $3\times 3$ Yukawa coupling matrix $Y_u^{bulk}$ is composed of ${\cal{O}}(1)$ dimensionless
coefficients. From these 5D interactions one can extract the 4D Yukawa
couplings (and the 4D mass terms) of the up-type quark zero modes 
(similarly for the down-type quark and the lepton sector of the theory). The
4D effective Yukawa couplings  between the zero mode Higgs and quarks are 
obtained from the overlap integrals of the quarks and Higgs wave functions along the
extra dimension.
The normalized  wave functions for the left-handed  doublet and the
right-handed singlet quarks are $q(y)=\sqrt{k} N_q e^{(2-c_q) k y}$ 
and $u(y)=\sqrt{k} N_u e^{(2+c_u) k y}$, respectively, where
$N_q=\sqrt{\frac{(2c_q-1)}{1-e^{(1-2c_q)ky_1)}}}$  and 
$N_u=\sqrt{\frac{(2c_u+1)}{-1+e^{(1+2c_u)ky_1)}}}$ are canonical
normalization factors\footnote{Note that, to simplify the notation, we omitted the flavour
indices.}.
The Yukawa couplings  then are
\bea
{\cal L}^{4D}_{Y}= Y_u^{bulk} \sqrt{k}  N_qN_u \int_0^{y_1} dy h_0(y) e^{-\Delta c ky}
+ h.c. ,
\label{4DYukawas}
\eea
where $\Delta c =c_q-c_u$.
The canonically normalized Higgs profiles are given by:
%and for each case we have the canonically normalized wave functions
\bea
{h_0}_{usual}(y)=  \sqrt{\frac{2(a-1)k}{(e^{2(a-1) ky_1} -1)}} e^{aky}\, ,
\eea
and
\bea
{h_0}_{odd}(y) =\sqrt{k} \frac{2ky_1}{\pi} e^{-ky_1}   e^{2ky} \sin{(b k y)}\ (1+ {\cal O}(b^2)) \,,
\eea
where in the case of the odd Higgs we
have neglected terms terms of order $b^2\lesssim 10^{-2}$ (but we keep
all terms for numerical calculations).
We can now calculate the  4D Yukawa couplings in each
Higgs regime:
\bea
y^{usual}_u =\frac{\sqrt{2}}{(2-\Delta c)}\  Y_u^{bulk} N_q N_u e^{(1-\Delta{c})ky_1}  
\eea
and
\bea
y_u^{odd}  =\frac{2}{(2-\Delta c)^2}\  Y_u^{bulk} N_q N_u e^{(1-\Delta{c})ky_1} .
\eea
The two results are surprisingly similar, that is, both have the same
exponential dependence on $\Delta c=c_q-c_u$. 
The ratio of these expressions is %an ${\cal O}(1)$ factor 
given by $\frac{\sqrt{2}}{2-\Delta c}$, which is essentially ${\cal O}(1)$,
within the usual range of the $c$-values\footnote{
This usual range is such that if the quark flavour structure is explained by the exponential
factors in $\Delta c$,  typical values are roughly found
between $\Delta c \sim 1.2$ (lightest quarks) and $\Delta c \sim 0$
(top quark).}, 
and thus the odd
Higgs regime addresses the flavour puzzle of the SM in the same way as
the usual bulk Higgs does, {\it i.e.}, through small masses and hierarchical mixing angles
which are a reflection of the geographical location of the fermion wave functions along the 
fifth dimension.

%

%
%%%%%%%%%%%%%%%%%%%%%%%%%%%%%%%%%%%%
\section{Existence, stability and naturalness}
\label{sec:stability}

We first consider the stability conditions for the trivial Higgs VEV, {\it i.e.} 
$v=0$. In this case there is no EWSB, but we
can still study the spectrum of Higgs perturbations around the $v=0$
vacuum. The Higgs KK modes in the unbroken phase still satisfy the same
equation as Eq. (\ref{h_eq})
%\bea
%h''_n - 4k h'_n - \mu_B^2 h_n + m_n^2 e^{2 ky}h_n=0
%\label{h_eq}
%\label{trivial}
%\eea
but now the boundary condition, Eq. (\ref{h_bc}), is modified:
\bea
%& h_n'(0) &=  M_0\ h_n(0) \\
& h_n'(y_1) &=  - M_1 h_n(y_1)\, .
\eea
We would like to know what is the most negative value for
$\mu_B^2$,  before the lightest Higgs mode, $h_0$, becomes tachyonic. The
threshold condition will be reached when the lowest mode $h_0$ is
massless, {\it i.e.} $m^2_0=0$. Moreover, it is known that the largest eigenvalue of the Sturm-Liouville
boundary value problem is  the eigenvalue
associated to the Dirichlet problem, which in this case is obtained in
the limit of very large $M_0$ and $M_1$. Therefore, the absolute stability
bound will be obtained for the value of $\mu_B^2$ that generates
a massless zero-mode ($m^2_0=0$) in the Dirichlet boundary value
problem associated with Eq. (\ref{h_eq}).  
The massless mode profile solution with Dirichlet boundary conditions is
${h_0}_{trivial}(y) = N_h e^{2 ky} \sin{bky}$, with the parameter $b$
fixed by the Dirichlet boundary condition as $b=\pi/(ky_1)$.
But the parameter $b$ depends on the bulk Higgs mass, $b^2k^2=-\mu_B^2 - 4k^2$, and
therefore the threshold value of the bulk mass parameter is
$(\mu_B^2)_{min}=-4k^2(1+\pi^2/(2ky_1)^2)$. Below this value, the Dirichlet Higgs perturbations
around the trivial background are unstable. This value represents the
generalized Breitenlohner-Freedman bound \cite{Breitenlohner:1982jf} for a scalar field theory defined on a slice of
$AdS_5$ \cite{Toharia:2008ug} and is exactly the same as the stability
bound for Higgs perturbations around the 
non-trivial odd Higgs VEV, mentioned earlier. Before
returning to the EWSB phase, we  discuss the more general case
in which the boundary conditions are not Dirichlet.
%It can be proved {\color{red} maybe a reference here, or something
%  more \color{blue} no reference that I know... This is ours I
%  think. Maybe Quiros comments on it? (it does not seem though)} that
The (perturbative) expression for the mass of 
the lowest Higgs KK perturbation around the trivial vacuum, $v=0$, has
the same form as the mass of the lightest Higgs mode with
the non-trivial VEV. The only difference is that it has opposite sign
and is divided by a factor of 2, {\it i.e.}
\bea
{m_0^2}_{trivial} = - \frac{1}{2} {m_0^2}_{non-trivial}\,,
\eea
where the expressions for ${m_0^2}_{non-trivial}$ were given in Eqs. (\ref{mbulk}) and
(\ref{modd}), depending on the regime of the
nontrivial VEV.
% {\it i.e.} depending on 
%whether $\mu_B^2>-4k^2$ (usual bulk Higgs) or $\mu_B^2 < -4k^2$ (odd Higgs).
This means that whenever the unbroken phase is
unstable, the broken phase will be stable and vice versa.

We now return to the case where the Higgs acquires a nontrivial
VEV. In \cite{Toharia:2010ex} it was shown that for any bulk scalar potential
defined on a slice of $AdS_5$,  with the scalar field respecting
Dirichlet boundary conditions, any non-trivial scalar VEV solution with nodes
in the interval would be unstable. In the case of the simple potential
used in this paper, this result can be obtained 
easily, since the equation for the perturbations around a nontrivial
VEV is the same as the one for the trivial vacuum. Indeed, one finds that
the threshold at which a massless mode appears is precisely the point
where the VEV has a node at the boundaries (it is Dirichlet-like).
When the boundary node moves slightly into the
bulk, the eigenvalue becomes slightly negative. For
the case of a more general bulk Higgs potential, 
this no-node criterion continues to be useful in the determination
of the absolute stability bound associated with the Higgs sector.

A related topic is the size of the Higgs potential
parameters, namely the bulk mass coefficient, $\mu_B^2$, the brane
quadratic coefficients, $M_0$ and $M_1$, and 
the brane quartic coefficient, $\gamma_1$.
As previously discussed, the parameter $M_0$ can
remain to be ${\cal O} (1)$ (in units of $k$) without changing any 
properties of the odd Higgs VEV
solution, Eq. (\ref{eq:vodd}).
In the case of  the bulk mass parameter, it is bound to be 
 $\mu_B^2 > -4k^2-\pi^2/(ky_1)^2k^2$ and so it can easily remain ${\cal O} (1)$ in units of
$k^2$.

Moreover, as we have seen earlier, in order to have a light Higgs the
bulk mass parameter must be very close to the stability limit, i.e. close to $-4k^2$.
This implies some degree of parameter tuning because one needs to
require the function $F(b)$ to be small enough to generate the light
Higgs mass (see Eq. (\ref{modd})). 
If we define
\bea
\epsilon_b=\mu_B^2+4k^2+\pi^2/y_1^2,
\label{epsilonb}
\eea
assuming small $\epsilon_b$, the approximate expression for
$F(b)$ (including the dependence on $M_1$) becomes
\bea
F(b) \simeq \frac{\epsilon_b ky_1}{b}
\left((4+ M_1/ k ) \frac{\epsilon_b ky_1}{b}  - 1 \right),
\eea
%where the definition of $b$ is again $b^2=-\mu_B^2/k^2-4$.
from which we can easily identify  two contributions to the overall coefficient.
%    that in the
%    parameter space region where the electroweak parameters are
%    suppressed.
Since $(ky_1)\sim 34$ and $b\sim 10^{-1}$, one needs at least
$\epsilon_b\sim 10^{-3}$ so that: % the overall factor
$\left(\frac{\epsilon_b ky_1}{b}\right)  \sim 10^{-1}$. This
requirement on $\epsilon_b$ represents roughly a $0.1\%$ tuning on the parameter
$\mu_B^2$ and we still need a further $10^{-1}$ factor from the term
containing $M_1$ in order to obtain the overall $10^{-2}$ suppression for $F(b)$.
We see that, while in the normal bulk Higgs case (Eq. (17)),  $M_1$ has to be
  negative, in this scenario it can be positive or negative. In fact,
  with our previous choice, we see that  a positive value of
  $M_1/k$ lying between $4$ and  $5$ can achieve the overall
  suppression of $F(b)$ without really much tuning of $M_1$. We
  conclude that in this regime, the required tuning to obtain a light
  Higgs is around $0.1\%$ and the tuned parameter is $(\mu_B^2)$.\footnote{  
%\red{? seams like here $M_1$ doesn't need to be negative, and also we should be good if say $M_1$ is 4k or 5k, so not tuned at all... then $0.1 ( (4 + 4 or 5 ) 0.1 - 1) \sim -0.01 or -0.02 \sim ( {weak scale \over \lambda_{KK}} )^2 \sim 0.01$ } {\color{blue} I get $(M_1/k-6) \sim O(10^{-1}-1)$), meaning $M_1/k \sim 6-7$, and I guess this is tuned, even if it is not small or negative. Does this not conflict with Eq. (17) which requires $M_1$ to be negative?}
%{\color{green}Not sure where the $(M_1/k-6)$ appears. The equation for $F(b)$ is
%  equation (19) and expanding it I find the form given in the text
%  with $(4+M_1/k)$. In the normal bulk Higgs case (Eq (17))  $M_1$ has to be
%  negative. In this scenario it can also be positive (can be both)
%  and in the interesting %region it is postive (somewhere between 4/k and 15/k). }
%% We conclude that essentially
%%   This leaves %So overall we have to fix %separately both $M_1$ and
%% $\mu_B^2$ still to be fixed, with an overall precision (tuning) of the
%% order of ($10^{-3}$) which can be viewed as a drawback of the odd
%% Higgs regime proposed in this letter.
Note that, as we mentioned before, ({\it cf.} Eq. (\ref{mbulk}) and
the discussion below it), in the usual bulk Higgs scenario there is
also a tuning,  
%which is
%explicitly written between the
%parameters $a$ and $M_1$ 
%but it can be reformulated as 
$(M_1+2k+\sqrt{\mu_B^2+4k^2}) \sim {\cal O}(1\%)$. %(10^{-2}$-$10^{-3}$).
This tuning is initially slightly less 'fine' than in
the odd Higgs case for a given  KK scale. However, since the odd Higgs
allows for a lower KK scale, the tuning ends up becoming of similar
order if one considers a bulk Higgs scenario with heavier KK modes
with masses closer to 10 TeV.}
%, {\it i.e.}, a little hierarchy emerges in both cases.}

The brane quartic coefficient $\gamma_1$ is
 fixed by the requirement of obtaining a light enough electroweak
scale, {\it i.e.} generating the appropriate $Z$ and $W$ boson masses.
The $W$ mass can be approximated as
\bea
m_W^2 = \frac{g_5^2}{4y_1} \int_0^{y_1} dy v^2(y) e^{-2ky}
\label{Wmass}
\eea
where $v(y)$ is the nontrivial Higgs VEV, which depending on the sign
of $(4k^2+\mu^2)$, can take either the usual form of Eq.~(\ref{vusual}) or
the new oscillatory solution $v_{odd}(y)$ of Eq.~(\ref{v_odd}).

The constant coefficient $v_0$ can be removed in favor of $m_W^2$, and
using the IR boundary condition for the VEV, Eq.~(\ref{vbc}), we can solve
for the quartic coefficient, $\gamma_1$, as
\bea
\gamma_1=\frac{g_4^2}{16}\frac{m_h^2}{m_W^2} \left(\int^{y_1}_0 dy
\frac{e^{-2 k y} v(y)^2}{e^{-2 k y_1}v(y_1)^2}    \right)^2.
\eea
When $v(y)=v_{usual}(y)$, one obtains $\gamma_1
=\frac{g_4^2}{16k^2}\frac{m_h^2}{m_W^2} \frac{1}{(2a-2)^2}$ which is
naturally of ${\cal O}(1/k^2)$.
However when $v(y)=v_{odd}(y)$ the coefficient $\gamma_1$ diverges
if $v_{odd}$ vanishes at the IR brane ({\it i.e.} when it becomes exactly
Dirichlet-like). This is not surprising since Dirichlet boundary
conditions    %it simply reflects the fact that 
require an infinite
brane potential.  However, $v_{odd}(y)$ does not exactly vanish at the IR brane
as the Higgs mass should not be zero. The parameter $\epsilon_b $
as defined in Eq. (\ref{epsilonb}) represents the deviation from the stability threshold ({\it i.e.}, a massless
Higgs) and 
$\gamma_1$ can be easily obtained for different  values of $\epsilon_b$.  For example, for a fixed
volume factor $ky_1=34$ and 
$\epsilon_b=10^{-3}$ we obtain $\gamma_1\simeq \left(
\frac{0.86}{k^2}\right) $, while for $\epsilon_b=5\times 10^{-4}$ we get
$\gamma_1 \simeq \left(  \frac{7.1}{k^2}\right)$ and for
$\epsilon_b=10^{-4}$, $\gamma_1\simeq\left(  \frac{2.5\times
  10^3}{k^2}\right)$. 
This behavior reflects again %the fact 
that in order to suppress scalar
field values at the boundary, one requires a large scalar brane
potential. Moreover, we also see that the value of the quartic
potential acquires very large values close to the region where the
 oblique parameters are most strongly suppressed. However if one
insists in keeping $\gamma_1$ at most ${\cal O} (1/k^2)$ one can still
have a suppressed $T$ parameter for $\epsilon_b \sim  10^{-3}$ (see Figs. 1
and 2).

Finally we turn to the interplay between the Higgs background
and the gravitational background of this scenario. So far we have  assumed
a static RS background, but it is known that since the background
contains a massless graviscalar mode, the 
radion, it must be stabilized. The natural question  is whether the back-reaction of the
nontrivial Higgs VEV (which we  neglected, assuming it to be
small) will stabilize the gravitational background, namely if the
radion will acquire a positive mass squared.
Unfortunately, it was shown in \cite{Lesgourgues:2003mi} that when one considers static
solutions for both the warp factor and a single scalar field (here, the
Higgs), the radion will be tachyonic whenever the derivative of
the scalar VEV  vanishes inside the interval.
Since this is precisely what happens with the odd Higgs VEV,  the
setup as is will not be gravitationally stable. However, RS type scenarios
always require a mechanism to lift the massless radion. The simplest
procedure is to add a new scalar singlet which would acquire a VEV
and will back-react on the metric background generating a stabilizing potential for the radion
\cite{Goldberger:1999uk,Goldberger:1999un,Csaki:2000fc,Csaki:1999mp,DeWolfe:1999cp,Arkani-Hamed:1998cxo}. This
is the solution for our scenario as well, except that in our case the stabilizing scalar should lift
the radion tachyonic mass generated by the Higgs nontrivial VEV. Since
the Higgs VEV determines the electroweak scale (which is much smaller than the KK
scale), one expects the scale of the radion tachyonic mass %, which
%is generated by the back-reaction of the Higgs VEV on the metric,
to also be small, and thus raising the radion mass should proceed in a 
fashion similar to that  of the usual warped scenarios.

%%%%%%%%%%%%%%%%%%%%%%%%%%%%%%%%%%%%
\section{Conclusions}
We considered the simplest RS warped scenario in which matter particles and the
Higgs field propagate in the bulk and  where the flavour structure of
the SM  is generated by the localization of the fermion fields along the extra space dimension. 
We  found that a small region of parameter space within this setup, where the bulk Higgs mass is pushed beyond the naive
$-4k^2$ threshold,
has not been explored yet.  Moreover, within that region the bounds from
precision electroweak constraints are much milder than in the usual bulk
Higgs regimes. In this region, we carefully addressed stability concerns associated to the extension of the
parameter space.

Even though the regime considered here requires slightly more tuning than
 the usual bulk Higgs regime, and also can require a somewhat larger brane
 scalar potential, this scenario is very interesting thanks 
to its simplicity and its success in reducing the $T$ parameter.
Even if a small hierarchy problem is present with the setup, the
oscillatory Higgs VEV can be exploited within different scenarios, either with custodial
symmetry or with heavily modified metric backgrounds. Thus, 
while still addressing the hierarchy and
the flavour problem of the Standard Model, 
the bounds and mass limits of the model are improved, making it possible to allow for
lighter KK modes, which would be accessible at the LHC.

\section{Aknowledgements}

The work of M. F.  is partly supported by NSERC
through the grant number SAP105354.
M. T. would like to thank FRQNT for financial support under              
grant number PRC-290000. This paper reflects solely the authors'
personal opinions and does not represent the opinions of the authors'
employers, present and past, in any way.

\bibliographystyle{elsarticle-num-names} 
\bibliography{oddhiggs}

\end{document}